\documentclass{emulateapj}

\def\hda{\mbox{HD~2454}}
\def\hdb{\mbox{HD~16220}}
\def\hdc{\mbox{HD~76932}}
\def\hdd{\mbox{HD~94028}}
\def\hde{\mbox{HD~107113}}
\def\hdf{\mbox{HD~140283}}
\def\hdg{\mbox{HD~160617}}

\def\kmsec{\mbox{km~s$^{\rm -1}$}}
\def\logg{\mbox{log~{\it g}}}

\def\teff{\mbox{$T_{\rm eff}$}}
\def\vt{\mbox{$v_{\rm t}$}}
\def\rpro{\mbox{$r$-process}}
\def\ppro{\mbox{$p$-process}}
\def\spro{\mbox{$s$-process}}

\def\loggf{$\log$($gf$)}

\shorttitle{Ge, As, and Se in Metal-Poor Stars}
\shortauthors{I.~U.\ Roederer}
\slugcomment{Accepted for publication in the Astrophysical Journal}

\begin{document}

\title{
Germanium, Arsenic, and Selenium Abundances in Metal-Poor 
Stars\footnotemark[1]
}

\footnotetext[1]{
Based on observations made with the NASA/ESA 
\textit{Hubble Space Telescope},
obtained from the data archive at the Space Telescope Science Institute. 
STScI is operated by the Association of Universities for 
Research in Astronomy, Inc.\ under NASA contract NAS 5-26555.
This research made use of StarCAT, hosted by the Mikulski
Archive at the Space Telescope Science Institute (MAST). 
These data are associated with Programs
GO-7348,
GO-7433,
GO-8197,
GO-9048,
GO-9455, and 
GO-9804. \\
\indent
Based on data obtained from the European Southern Observatory (ESO) 
Science Archive Facility.
These data are associated with Programs
67.D-0439(A),
074.C-0364(A),
076.B-0055(A), and
080.D-0347(A). \\
\indent
This research has made use of the Keck Observatory Archive (KOA), 
which is operated by the W.M.\ Keck Observatory and the NASA Exoplanet 
Science Institute (NExScI), under contract with the 
National Aeronautics and Space Administration.
These data are associated with Programs
H2aH, H6aH, and H39aH (P.I.\ Boesgaard),
N01H (P.I.\ Latham), and
U11H (P.I.\ Prochaska). \\
\indent
This paper includes data taken at The McDonald Observatory of The 
University of Texas at Austin.
}

\author{
Ian U.\ Roederer}
\affil{Carnegie Observatories, 
813 Santa Barbara Street, Pasadena, CA 91101, USA}
\email{
iur@obs.carnegiescience.edu}


\addtocounter{footnote}{1}

\begin{abstract}

The elements
germanium (Ge, $Z =$~32), 
arsenic (As, $Z =$~33), and 
selenium (Se, $Z =$~34)
span the transition from charged-particle
or explosive synthesis of the iron-group elements
to neutron-capture synthesis of heavier elements.
Among these three elements,
only the chemical evolution of germanium has been 
studied previously.
Here we use archive observations made with the 
Space Telescope Imaging Spectrograph on board the
\textit{Hubble Space Telescope}
and observations from several ground-based facilities
to study the chemical enrichment histories of seven stars
with metallicities $-$2.6~$\leq$~[Fe/H]~$\leq -$0.4.
We perform a standard abundance analysis of germanium,
arsenic, selenium, and several other elements
produced by neutron-capture reactions.
When combined with previous derivations of germanium abundances in
metal-poor stars,
our sample reveals an increase in the [Ge/Fe] ratios
at higher metallicities.
This could mark the onset of the weak $s$-process
contribution to germanium.
In contrast, 
the [As/Fe] and [Se/Fe] ratios remain roughly constant.
These data do not directly indicate the origin of
germanium, arsenic, and selenium at low metallicity,
but they suggest that the weak and main components
of the $s$-process are not likely sources.

\end{abstract}

\keywords{
nuclear reactions, nucleosynthesis, abundances ---
stars: abundances ---
stars: individual (HD~2454, HD~16220, HD~76932, HD~94028, HD~107113, 
HD~140283, HD~160617)
}

\section{Introduction}
\label{intro}

An incomplete but nevertheless intriguing
history of stellar nucleosynthesis may be reconstructed
from the abundance patterns revealed by the heavy elements
present today.
The atmospheres of cool stars
largely preserve the initial chemical composition
of the interstellar medium (ISM)
at the time and place where each star formed.
Neutral and singly-ionized atoms, the majority species
in these atmospheres,
absorb radiation at energies in the optical 
spectrum accessible to ground-based telescopes.
These favorable conditions 
allow us to study the chemical record 
in the Sun and other stars
whose ages span the full age of the Universe.

We can interpret the origin of
elements that are readily detected in these stars
with the help of models of stellar structure and chemical evolution,
as well as steadily growing volumes of laboratory atomic and nuclear 
physics data.
For example,
substantial fractions of the iron-group elements in the Solar system 
were produced under explosive conditions in 
equilibrium or quasi-equilibrium
during previous generations of supernovae.
The overwhelming majority of the Solar composition of
significantly heavier elements, such as
barium (Ba, $Z =$~56) or europium (Eu, $Z =$~63),
must have been produced by neutron-capture
($n$,$\gamma$) reactions and subsequent $\beta^{-}$ decays.
Insurmountable Coulomb barriers 
prevent charged-particle reactions from 
contributing all but a negligible fraction of such elements.
Elements between these two extremes, such as
strontium (Sr, $Z =$~38) or
zirconium (Zr, $Z =$~40),
may be produced by each of these mechanisms.

Elements in this transition between zinc (Zn, $Z =$~30)
and rubidium (Rb, $Z =$~37)
are rarely studied in late-type stars because their 
intrinsic abundances are relatively low and their 
strongest absorption lines lie in the ultraviolet (UV), 
below the atmospheric cutoff near 3000\,\AA.
Some of these intermediate elements 
have been detected
in the atmospheres of so-called chemically-peculiar stars
(e.g., \citealt{bidelman62,leckrone99,castelli04,cowley10}).
These atmospheres are affected by strong
magnetic fields or are stable against convection,
allowing radiatively driven diffusion or gravitational settling
to chemically stratify the atmosphere.
These stars are not useful records for 
examining the evolution of specific elements
throughout the history of the Galaxy.
Gallium (Ga, $Z =$~31) has been detected in
just a few unstratified stars by \citet{ivans03,ivans06}.
Several high ionization states of some of these elements 
are found in planetary nebulae
(e.g., \citealt{pequignot94,dinerstein01,sterling02,sharpee07,sterling08}).
Recent work by \citet{karakas07,karakas09} interprets
these abundances in terms of self-enrichment 
by the slow neutron capture process (\spro).
This process operates in intermediate-mass stars during
the asymptotic giant branch (AGB) phase of evolution,
which precedes the planetary nebula stage.
Some of these elements are also detected in the ISM of the Milky Way
(e.g., \citealt{pwa86,cardelli91,hobbs93})
or other galaxies at high redshift
(e.g., \citealt{prochaska03}).
While general conclusions can be made,
it is difficult to probe 
stellar nucleosynthesis using the integrated heavy elements
found in this gas
at the level of detail afforded by individual stars in our own Galaxy.

Among the elements between zinc and rubidium,
only germanium (Ge, $Z =$~32) has been examined in more than
a handful of late-type stars
\citep{sneden98,sneden03,cowan02,cowan05,roederer12b}.
Yet even this sample comprises only old halo stars with metallicity
[Fe/H]~$< -$1.4, so the evolution of germanium
in younger disc stars with higher metallicity is still unknown.
Recently, \citet{roederer12c} detected neutral arsenic (As, $Z =$~33)
and selenium (Se, $Z =$~34) in the metal-poor halo star
\hdg\ using a high-resolution UV spectrum obtained
with the Space Telescope Imaging Spectrograph (STIS)
on board the \textit{Hubble Space Telescope} (\textit{HST}).
That study was able to place an interesting upper limit 
on the germanium abundance through the non-detection of
Ge~\textsc{i}.
With the exception of one good Ge~\textsc{i} absorption line at 3039\,\AA,
the only detectable transitions 
of these elements
lie near 2000\,\AA\ in the UV.
No ground-based telescope offers the opportunity
to study all three of these elements simultaneously.

We have searched the \textit{HST} archives to find any 
stellar spectra that offer 
a reasonable opportunity to detect the rarely-studied elements
germanium, arsenic, and selenium.
We find suitable spectra of six 
stars that span a range of metallicities
and stellar populations.
We summarize some relevant properties of these elements
in Section~\ref{solar},
characterize the archival observations 
in Section~\ref{observations},
describe our analysis in \ref{analysis},
present our results in Section~\ref{results}, and
summarize our conclusions in Section~\ref{conclusions}.

Throughout this work we use
standard definitions of elemental abundances and ratios.
For element X, the abundance is defined
as the logarithm of the 
number of atoms of element X per 10$^{12}$ hydrogen atoms,
$\log\epsilon$(X)~$\equiv \log_{10}(N_{\rm X}/N_{\rm H}) +12.0$.
For elements X and Y, the abundance ratio relative to the
solar ratio of X and Y is defined as the logarithm
[X/Y]~$\equiv \log_{10} (N_{\rm X}/N_{\rm Y}) -
\log_{10} (N_{\rm X}/N_{\rm Y})_{\odot}$.
Abundance ratios, e.g., [X/Fe], 
are constructed
by comparing the total abundance of element X derived from the neutral species 
with the total iron abundance derived
from Fe~\textsc{i} or the total abundance of element X derived from the
ionized species with the total iron abundance
derived from Fe~\textsc{ii}.
Abundances or ratios denoted with the ionization state
indicate the particular ionization state used to derive
the total number of atoms of the element under consideration.

\section{Germanium, Arsenic, and Selenium in the Solar System}
\label{solar}

Five isotopes of germanium occur naturally in the Solar system,
$^{70,72,73,74,76}$Ge.
The lightest four of these isotopes are accessible to the \spro,
and even $^{76}$Ge may be accessible to the
\spro\ if the branching point at $^{75}$Ge is open.
All but $^{70}$Ge are accessible to the rapid ($r$) neutron-capture process.
Only one isotope of arsenic is found in Solar material,
$^{75}$As, which may be produced by both $s$- and \rpro\ 
nucleosynthesis.
There are six naturally-occurring isotopes of selenium,
$^{74,76,77,78,80,82}$Se.
Four of these isotopes, 
$^{76,77,78,80}$Se, lie along the \spro\ chain, and
only $^{74}$Se and $^{76}$Se are shielded from the \rpro.
$^{74}$Se may be produced by 
proton-capture ($p$,$\gamma$) or other
reactions that produce nuclei on the proton-rich 
side of stability.
$^{74}$Se constitutes only a minor fraction (0.89\%) of the
total elemental abundance, and we shall not consider it further here.
The Solar isotopic abundances of these three elements are well-known
and are summarized in, e.g., \citet{bohlke05}.

Several of the heavier isotopes 
of selenium derive a large fraction 
of their Solar abundance from the \rpro\ production
of radioactive isotopes with 50~neutrons.
This stable nuclear configuration gives rise to smaller
than average cross sections to neutron capture reactions,
building up a peak in the \rpro\ distribution.
After the \rpro\ shuts off, these radioactive isotopes 
$\beta^{-}$ decay to stable isotopes of selenium.
Selenium is unique among the elements in our study because
it lies at the first \rpro\ peak.
Similar phenomena occur in nuclei with 82 and 126
neutrons, the second and third peaks, respectively.

The classical approach to calculating the \rpro\ 
contribution to the Solar abundance distribution,
$N_{r} = N_{\odot} - N_{s}$,
neglects to account for the multitude of 
nucleosynthesis channels that can produce 
elements between the iron-group and the first \spro\ peak
(krypton through zirconium, 36~$\leq Z \leq$~40).
These decompositions assigned a very large
\spro\ fraction ($\sim$20--70\%) to the 
Solar germanium, arsenic, and selenium
abundances (e.g., \citealt{kappeler82} or \citealt{burris00};
see \citealt{cameron82} for a similar approach).
Modern decompositions using the stellar model approach of
\citet{gallino98}, \citet{arlandini99}, and the updates
given in \citet{bisterzo11} assign only 5--10\% of the
Solar abundances of germanium, arsenic, and selenium
to the main component of the \spro.
The \ppro\ contributions to the Solar
abundances are expected to be quite small
(e.g., \citeauthor{kappeler89}).
The remainder is assumed to originate in the weak
component of the \spro\ or mechanisms 
associated with core-collapse supernovae.
\citet{pignatari10} estimate the primary, or explosive, component 
of the Solar germanium abundance to be 5--8\%.
Applying a model of Galactic chemical evolution, 
\citet{travaglio04a} estimate another 12\% of the Solar germanium
should originate in low-mass stars
passing through the AGB phase.
Type~Ia supernovae should produce no significant germanium
(e.g., \citealt{travaglio04b}).
The remaining 80\% of the Solar abundance of germanium 
should be produced by the weak \spro\ operating in massive 
stars (e.g., \citeauthor{pignatari10}).
Regardless of the exact fractions, modern methods of
accounting for the various nucleosynthesis contributions
make clear that the main \spro\ component had a 
relatively minor influence on the total Solar abundances
of these elements.

\section{Observations from the Archives}
\label{observations}

\begin{deluxetable*}{cccccccc}
\tablecaption{Characteristics of the Archival Spectra
\label{obstab}}
\tablewidth{0pt}
\tabletypesize{\scriptsize}
\tablehead{
\colhead{Star} &
\colhead{$\lambda$ (\AA)} &
\colhead{Instrument} &
\colhead{$t_{\rm exp}$ (s)} &
\colhead{$R \equiv \lambda/\Delta\lambda$} &
\colhead{S/N\,$@$\,$\lambda$} &
\colhead{Program} &
\colhead{P.I.} }
\startdata
\hda\ & 1970--2770 & STIS  &  1929 &  30,000 &  40\,$@$\,2000\,\AA\ & GO-9048       & Deliyannis \\
      & 2290--3110 & STIS  &  1440 &  30,000 &  50\,$@$\,3050\,\AA\ & GO-7433       & Heap \\
      & 3030--4515 & HIRES &   420 &  49,000 &  50\,$@$\,3500\,\AA\ & H2aH          & Boesgaard \\
      & 3790--6905 & HARPS &    60 & 115,000 & 100\,$@$\,5000\,\AA\ & 074.C-0364(A) & Robichon \\
      & 5755--8150 & HIRES &   120 &  49,000 & 500\,$@$\,6000\,\AA\ & H39aH         & Boesgaard \\
\hdb\ & 1990--2815 & STIS  &  1975 &  30,000 &  30\,$@$\,2100\,\AA\ & GO-9048       & Deliyannis \\
      & 3050--3825 & HIRES &   100 &  49,000 & 150\,$@$\,3500\,\AA\ & U11H          & Prochaska \\
      & 3950--4905 & HIRES &   100 &  49,000 & 300\,$@$\,4500\,\AA\ & U11H          & Prochaska \\
      & 4950--5855 & HIRES &   100 &  49,000 & 200\,$@$\,5500\,\AA\ & U11H          & Prochaska \\
\hdc\ & 1880--2145 & STIS  & 23859 & 110,000 &  30\,$@$\,2000\,\AA\ & GO-9804       & Duncan \\
      & 3050--3865 & UVES  &   900 &  51,000 & 250\,$@$\,3500\,\AA\ &  67.D-0439(A) & Primas \\
      & 3760--4980 & UVES  &    96 &  41,000 & 150\,$@$\,4000\,\AA\ & 076.B-0055(A) & Silva \\
      & 4785--5750 & UVES  &   130 &  62,000 & 350\,$@$\,5000\,\AA\ &  67.D-0439(A) & Primas \\
      & 5835--6805 & UVES  &    30 &  62,000 & 300\,$@$\,6000\,\AA\ &  67.D-0439(A) & Primas \\
\hdd\ & 1885--2145 & STIS  & 33048 & 110,000 &  20\,$@$\,2000\,\AA\ & GO-8197       & Duncan \\
      & 3935--4810\tablenotemark{a} 
                   & HIRES &   270 &  67,000 & 200\,$@$\,4500\,\AA\ & N01H          & Latham \\
      & 3695--$\approx$7700& Tull  &  3300 &  33,000 & 145\,$@$\,5200\,\AA\ & New observation & Roederer \\
\hde\ & $\approx$1800--2360 & STIS  &  2281 &  30,000 &  40\,$@$\,2000\,\AA\ & GO-7433       & Heap \\
      & 4430--6770 & HIRES &   180 &  50,000 & 350\,$@$\,5000\,\AA\ & H2aH          & Boesgaard \\
\hdf\ & 1935--2210 & STIS  &  8337 & 110,000 &  25\,$@$\,2000\,\AA\ & GO-7348       & Edvardsson \\
      & 2390--3135 & STIS  & 11077 &  50,000 &  40\,$@$\,3000\,\AA\ & GO-9455       & Peterson \\
      & 3785--6915 & HARPS &  2700 & 115,000 & 250\,$@$\,4500\,\AA\ & 080.D-0347(A) & Heiter \\
\enddata
\tablenotetext{a}{The iodine cell was inserted during these observations, 
so we only make use of the blue region of the spectrum unaffected by
I$_{2}$ absorption.}
\end{deluxetable*}

We have searched the \textit{HST} archives 
to find high-resolution 
spectral observations covering
wavelengths from 1890~$\leq \lambda \leq$~2074\,\AA\
where Ge~\textsc{i}, As~\textsc{i}, and Se~\textsc{i}
lines may be found.
The archive contains
six subgiants with adequate signal-to-noise (S/N) ratios
(S/N $\gtrsim$~20 near 2000\,\AA)
and similar stellar parameters:
\hda, \hdb, \hdc, \hdd, \hde, and \hdf.
\citet{roederer12c} also studied
these elements in an archival STIS spectrum
of another star, \hdg.
These stars comprise our target sample
for the present study.

All UV spectra were taken with STIS
\citep{kimble98,woodgate98}.
These spectra were obtained using either the
high resolution E230H grating or the 
medium resolution E230M grating.
Table~\ref{obstab} reports the details of each
set of observations, including
wavelength coverage, exposure time, approximate
spectral resolution,
S/N ratios at a few reference wavelengths,
program identifier, and the principle investigator (P.I.)
of the original observations.
We use the reduction and coaddition provided by the
StarCAT database \citep{ayres10}.

These are local (20~$< d <$~60~pc), 
bright stars (5.8~$< V <$~8.7), and they
have been observed many times using ground-based instruments.
We supplement the UV spectra
with optical spectra obtained with the 
High Accuracy Radial velocity Planet Searcher (HARPS; \citealt{mayor03}) 
on the ESO 3.6~m Telescope at La Silla, Chile; the
Ultraviolet and Visual Echelle Spectrograph (UVES; \citealt{dekker00}) 
on the Very Large Telescope (VLT) Kueyen at Cerro Paranal, Chile; and the
High Resolution Echelle Spectrometer (HIRES; \citealt{vogt94})
on the Keck~I Telescope at Mauna Kea, Hawai'i.
These spectra were reduced and extracted
using the standard instrument pipelines, and we performed the
final processing within the IRAF environment.
New observations of \hdd\ were obtained
on 2009 February 18 and 2009 June 12
using the Robert G.\ Tull Coud\'{e} Spectrograph
\citep{tull95} on the 2.7~m Harlan J.\ Smith Telescope 
at McDonald Observatory, Texas.
These data were reduced and combined using the REDUCE software
package \citep{piskunov02} and standard IRAF procedures.
We use this spectrum to supplement the HIRES spectrum of \hdd\
at longer wavelengths.
The details of each of these observations are also listed in
Table~\ref{obstab}.
Our previous study of \hdg\ revealed no
systematic differences at the 1~m\AA\ level
between sets of equivalent widths (EWs)
measured from high-quality spectra collected with these instruments.

\section{Abundance Analysis}
\label{analysis}

\subsection{Model Atmospheres}

\begin{deluxetable*}{ccccccccc}
\tablecaption{Model Atmosphere Parameters and Iron Abundances
\label{atmtab}}
\tablewidth{0pt}
\tabletypesize{\scriptsize}
\tablehead{
\colhead{Star} &
\colhead{\teff} &
\colhead{\logg} &
\colhead{\vt} &
\colhead{[Fe~\textsc{i}/H]~$\pm$~$\sigma$} &
\colhead{$N$} &
\colhead{[Fe~\textsc{ii}/H]~$\pm$~$\sigma$} &
\colhead{$N$} \\
\colhead{} &
\colhead{(K)} &
\colhead{} &
\colhead{(\kmsec)} &
\colhead{} &
\colhead{} &
\colhead{} &
\colhead{} }
\startdata
\hda\ & 6050 & 4.04 & 1.20 & $-$0.72~$\pm$~0.10 & 69  & $-$0.44~$\pm$~0.21 & 7  \\
\hdb\ & 6000 & 3.81 & 1.30 & $-$0.56~$\pm$~0.14 & 51  & $-$0.41~$\pm$~0.22 & 7  \\
\hdc\ & 5680 & 4.11 & 1.00 & $-$1.18~$\pm$~0.12 & 112 & $-$0.92~$\pm$~0.12 & 11 \\
\hdd\ & 5720 & 4.31 & 0.90 & $-$1.74~$\pm$~0.09 & 44  & $-$1.62~$\pm$~0.05 & 6  \\
\hde\ & 6060 & 3.99 & 1.25 & $-$0.79~$\pm$~0.08 & 58  & $-$0.53~$\pm$~0.07 & 5  \\
\hdf\ & 5600 & 3.66 & 1.15 & $-$2.71~$\pm$~0.05 & 111 & $-$2.62~$\pm$~0.07 & 31 \\
\enddata
\tablecomments{We set the overall metal abundance of the
model equal to [Fe~\textsc{ii}/H].}
\end{deluxetable*}

We interpolate model atmospheres from the ATLAS9 grid of
\citet{castelli03}, using the $\alpha$-enhanced grid for 
stars with overall metallicity
[Fe/H]~$< -$0.7.
We use the latest version of the
analysis code MOOG \citep{sneden73}, including updates
from \citet{sobeck11}, to perform a standard EW abundance analysis
for iron.
We measure EWs of Fe~\textsc{i} and Fe~\textsc{ii} lines
using a semi-automated routine that fits Voigt absorption
line profiles to continuum-normalized spectra.
We adopt the \loggf\ values for Fe~\textsc{i} and Fe~\textsc{ii} lines
from the critical compilation of \citet{fuhr06}.
We adopt damping constants from \citet{barklem00}
and \citet{barklem05}, when available, and otherwise
we resort to the standard \citet{unsold55} approximation.
We calculate surface gravities, \logg\ (cm\,s$^{-2}$), based on the 
\textit{Hipparcos} parallax measurements validated by
\citet{vanleeuwen07}, bolometric corrections from \citet{alonso99},
stellar masses from \citet{casagrande11}, and Solar parameters
$M_{\rm bol} =$~4.74, 
\logg$_{\odot} =$~4.44, and
\teff$_{\odot} =$~5780~K.
We derive effective temperatures, \teff, by requiring that the 
abundances derived from individual Fe~\textsc{i} lines show
no dependence on lower excitation potential (E.P.).
We derive microturbulent velocities, \vt, by requiring that
these abundances show no trend with line strength.
We set the overall metallicity of the model atmosphere equal to the
iron abundance derived from Fe~\textsc{ii}.
Reddening along the line of sight to these stars is insignificant
and can be neglected 
(e.g., \citeauthor{casagrande11}).
We iterate on the model atmosphere parameters until reaching
convergence, which typically requires 2--4~iterations.

Our derived model atmosphere parameters and iron abundances
are listed in Table~\ref{atmtab}.
Uncertainties in the atmospheric parameters should be
comparable to those discussed in Section~4.1 of \citet{roederer12c}.
Our derived temperatures are consistently several hundred K cooler than
temperatures calculated using color-temperature relations or the
infrared flux method.
The warmer temperature scales all introduce a correlation
between E.P.\ and abundance, signaling the need for a 
cooler temperature scale.
This is not ideal, but other approaches (e.g., Balmer line profiles,
color-temperature relations) also incur difficulties.
Absolute abundances should be
viewed with caution.
Since all of our stars are subgiants or turn-off stars
with similar temperatures,
the abundance ratios of one heavy element to another should
be affected minimally.

\subsection{Derivation of Abundances}

\begin{deluxetable*}{ccccccccc}
\tablecaption{Line-by-Line Abundances
\label{linetab}}
\tablewidth{0pt}
\tabletypesize{\scriptsize}
\tablehead{
\colhead{Species} &
\colhead{$\lambda$} &
\colhead{\loggf} &
\colhead{HD 2454} &
\colhead{HD 16220} &
\colhead{HD 76932} &
\colhead{HD 94028} &
\colhead{HD 107113} &
\colhead{HD 140283} \\
\colhead{} &
\colhead{(\AA)} &
\colhead{} &
\colhead{$\log \epsilon$} &
\colhead{$\log \epsilon$} &
\colhead{$\log \epsilon$} &
\colhead{$\log \epsilon$} &
\colhead{$\log \epsilon$} &
\colhead{$\log \epsilon$} }
\startdata
Zn~\textsc{i} & 3302.58 & $-$0.02 & \nodata   & \nodata   & $+$3.56   & \nodata   & \nodata   & \nodata \\
              & 4680.14 & $-$0.85 & $+$3.96   & $+$4.13   & $+$3.71   & $+$3.06   & $+$3.88   & \nodata \\
              & 4722.16 & $-$0.37 & $+$3.91   & $+$4.10   & $+$3.76   & $+$3.10   & $+$3.81   & $+$2.17 \\
              & 4810.53 & $-$0.15 & $+$4.03   & $+$4.11   & $+$3.79   & $+$3.11   & $+$3.80   & $+$2.09 \\
Zn~\textsc{ii}& 2062.00 & $-$0.29 & \nodata   & \nodata   & \nodata   & $+$3.26   & \nodata   & $+$2.27 \\
Ge~\textsc{i} & 1998.89 & $-$0.78 & \nodata   & \nodata   & \nodata   & $+$1.63:  & \nodata   & $<+$1.1 \\
              & 2041.71 & $-$0.70 & \nodata   & $+$3.11:: & $+$2.40   & $+$1.59   & $+$2.54:  & $<+$0.6 \\
              & 2065.21 & $-$0.79 & \nodata   & \nodata   & \nodata   & $+$1.39:  & \nodata   & $<+$0.8 \\
              & 3039.07 & $+$0.07 & $+$2.70\tablenotemark{a} & \nodata & \nodata & \nodata & \nodata & $<+$0.6 \\
As~\textsc{i} & 1972.62 & $-$0.63 & $<+$3.9   & \nodata   & $+$1.70:: & \nodata   & $+$1.87:: & $+$0.09:\\
              & 1990.36 & $-$0.28 & $<+$2.9   & $<+$3.1   & \nodata   & $+$1.21:  & \nodata   & \nodata \\
Se~\textsc{i} & 1960.89 & $-$0.43 & \nodata   & \nodata   & $+$2.37:: & $+$2.18:  & \nodata   & $+$0.59:\\
              & 2039.84 & $-$0.74 & \nodata   & \nodata   & \nodata   & $+$2.03:  & \nodata   & $+$0.41 \\
              & 2074.78 & $-$2.26 & $<+$3.3   & $+$2.83:: & $+$2.45:  & $+$2.03   & $+$2.77:  & $+$0.79 \\
Sr~\textsc{ii}& 4077.71 & $+$0.15 & \nodata   & \nodata   & \nodata   & $+$1.50   & \nodata   & $+$0.13 \\
              & 4215.52 & $-$0.17 & \nodata   & \nodata   & \nodata   & $+$1.52   & \nodata   & $+$0.05 \\
Zr~\textsc{ii}& 1996.74 & $-$0.04 & \nodata   & \nodata   & $+$1.89   & $+$1.36   & \nodata   & \nodata \\
              & 3095.07 & $-$0.84 & \nodata   & $+$2.27   & $+$1.84   & \nodata   & \nodata   & \nodata \\
              & 3129.76 & $-$0.54 & \nodata   & \nodata   & $+$1.99   & \nodata   & \nodata   & \nodata \\
              & 3273.05 & $+$0.30 & $+$2.69   & $+$2.40   & $+$2.02   & \nodata   & \nodata   & \nodata \\
              & 3284.71 & $-$0.37 & \nodata   & \nodata   & $+$1.97   & \nodata   & \nodata   & \nodata \\
              & 3479.39 & $+$0.18 & \nodata   & \nodata   & $+$1.93   & \nodata   & \nodata   & \nodata \\
              & 3505.67 & $-$0.39 & \nodata   & $+$2.29   & $+$1.95   & \nodata   & \nodata   & \nodata \\
              & 3551.95 & $-$0.36 & \nodata   & $+$2.36   & $+$2.00   & \nodata   & \nodata   & \nodata \\
              & 3998.96 & $-$0.52 & \nodata   & \nodata   & \nodata   & $+$1.44   & \nodata   & $-$0.13 \\
              & 4050.32 & $-$1.06 & $+$2.64   & $+$2.29   & $+$1.92   & $+$1.46   & \nodata   & \nodata \\
              & 4149.20 & $-$0.04 & \nodata   & \nodata   & \nodata   & $+$1.53   & \nodata   & $-$0.22 \\
              & 4161.20 & $-$0.59 & \nodata   & \nodata   & $+$1.98   & $+$1.51   & \nodata   & \nodata \\
              & 4208.98 & $-$0.51 & $+$2.68   & $+$2.35   & $+$1.96   & $+$1.48   & \nodata   & $-$0.09 \\
              & 5112.27 & $-$0.85 & \nodata   & \nodata   & \nodata   & \nodata   & $+$2.06   & \nodata \\
Ba~\textsc{ii}& 4130.65 & $+$0.52 & \nodata   & $+$1.67   & $+$0.95   & \nodata   & \nodata   & \nodata \\
              & 4554.03 & $+$0.14 & $+$2.51   & $+$2.23   & \nodata   & $+$0.61   & $+$1.94:  & $-$1.37 \\
              & 5853.68 & $-$0.91 & $+$2.64   & $+$2.07   & $+$1.29   & $+$0.76   & $+$1.79   & \nodata \\
              & 6141.71 & $-$0.03 & \nodata   & \nodata   & \nodata   & \nodata   & \nodata   & $-$1.45 \\
              & 6496.90 & $-$0.41 & \nodata   & \nodata   & \nodata   & \nodata   & \nodata   & $-$1.29 \\
La~\textsc{ii}& 3988.51 & $+$0.21 & $+$1.05   & $+$0.82   & $+$0.22   & $-$0.24   & \nodata   & $<-$1.0 \\
              & 3995.74 & $-$0.06 & $+$1.06   & $+$0.80   & $+$0.21   & $-$0.34   & \nodata   & $<-$1.1 \\
              & 4086.71 & $-$0.07 & $+$1.23   & $+$0.77   & $+$0.27   & $-$0.28   & \nodata   & $<-$1.4 \\
              & 4123.22 & $+$0.13 & $+$1.11   & $+$0.88   & $+$0.28   & $-$0.25   & \nodata   & $<-$1.2 \\
              & 4662.50 & $-$1.24 & \nodata   & \nodata   & \nodata   & \nodata   & $+$0.63   & \nodata \\
              & 5114.56 & $-$1.03 & \nodata   & \nodata   & \nodata   & \nodata   & $+$0.72   & \nodata \\
              & 6262.29 & $-$1.22 & \nodata   & \nodata   & \nodata   & \nodata   & $+$0.67:  & \nodata \\
              & 6390.48 & $-$1.41 & \nodata   & \nodata   & \nodata   & \nodata   & $+$0.62:  & \nodata \\
Eu~\textsc{ii}& 3819.67 & $+$0.51 & \nodata   & \nodata   & \nodata   & \nodata   & \nodata   & $<-$1.7 \\
              & 3907.11 & $+$0.17 & \nodata   & \nodata   & $-$0.11   & \nodata   & \nodata   & $<-$1.4 \\
              & 4129.72 & $+$0.22 & $+$0.04   & $+$0.14   & $-$0.17   & $-$0.88   & \nodata   & $<-$1.7 \\
              & 6645.10 & $+$0.12 & \nodata   & \nodata   & $-$0.11   & \nodata   & $<+$0.4   & \nodata \\
Pt~\textsc{i} & 2049.39 & $+$0.02 & \nodata   & \nodata   & \nodata   & $+$0.20   & \nodata   & $<-$0.6 \\
Pb~\textsc{i} & 4057.81 & $-$0.22 & $+$2.08   & \nodata   & $+$1.28   & $<+$1.3   & \nodata   & $<+$0.7 \\
\enddata
\tablenotetext{a}{
Both the STIS and HIRES spectra yield equivalent abundances
for this line
}
\end{deluxetable*}

\begin{figure}
\begin{center}
\includegraphics[angle=0,width=3.25in]{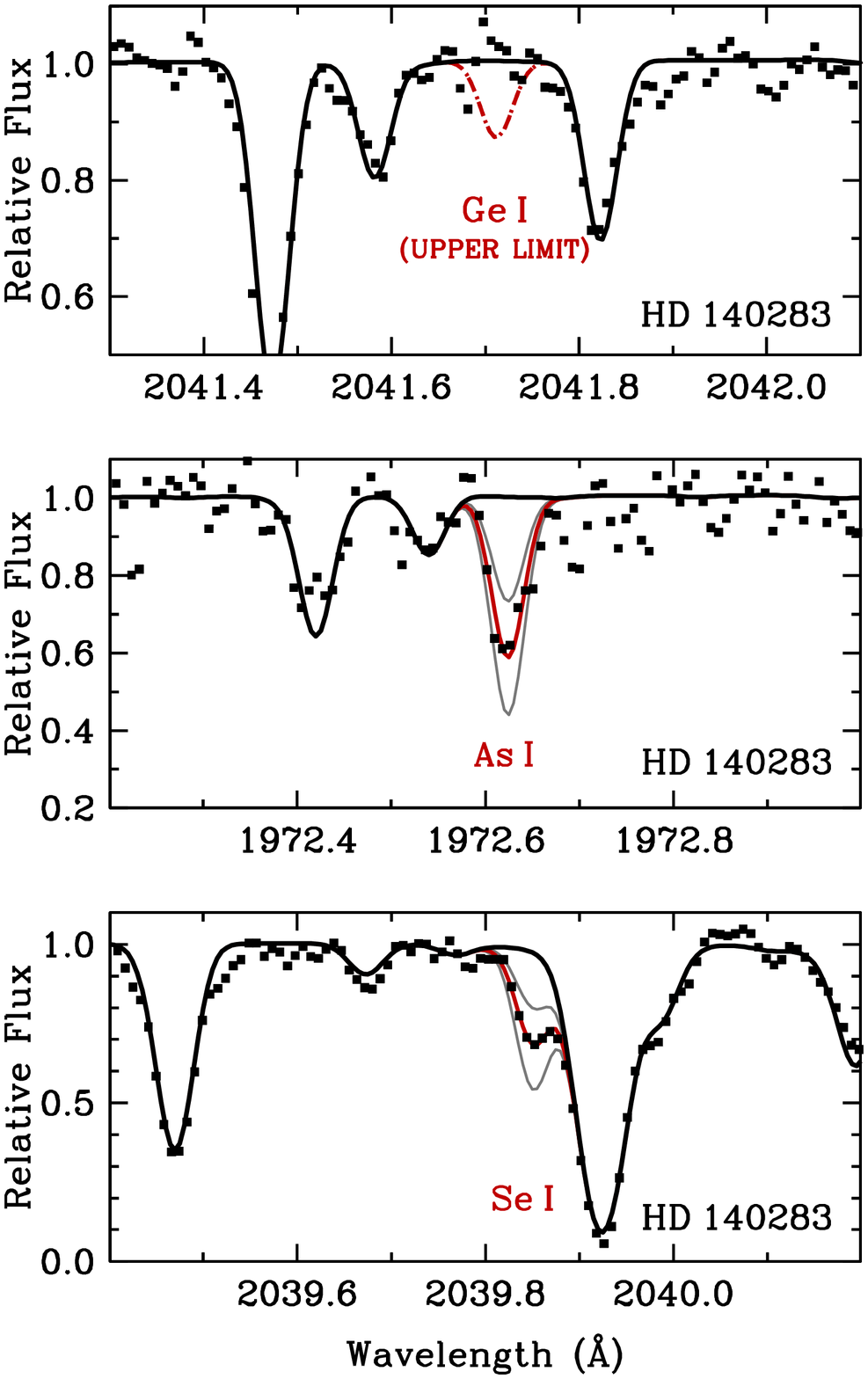}
\end{center}
\caption{
\label{synplot}
Synthesis of a sample of Ge~\textsc{i}, As~\textsc{i}, and Se~\textsc{i}
lines in the UV STIS spectrum of \hdf.
In the top panel, the Ge~\textsc{i} upper limit 
(3$\sigma$) is indicated by the
dot-dashed line.
In the lower two panels,
the best-fit synthesis is indicated by the bold red line.
The light gray lines show variations in the best-fit abundance by
$\pm$~0.3~dex.
In all three panels, the bold black line indicates a synthesis
with no Ge~\textsc{i}, As~\textsc{i}, or Se~\textsc{i} present.
 }
\end{figure}

Our techniques are the same as those employed by \citet{roederer12c}.
We derive all heavy element abundances by comparing 
synthetic spectra, computed using MOOG, with the observed spectrum.
Syntheses of several lines in \hdf\ are shown in Figure~\ref{synplot}
as examples.
Lines of other species near those of interest are synthesized based on the
\citet{kurucz95} line lists, updating their \loggf\ values
with more recent laboratory or theoretical values when available.
These lines have already been discussed by \citeauthor{roederer12c}.
In addition to germanium, arsenic, and selenium, we
also derive abundances of a selection of heavier elements
useful for interpreting the enrichment history of these stars:
zinc, 
strontium, 
zirconium, 
barium, 
lanthanum (La, $Z =$~57), 
europium, 
platinum (Pt, $Z =$~78), and
lead (Pb, $Z =$~82).
In all cases, we adopt the Solar abundances from \citet{asplund09}:
iron, $\log \epsilon =$~7.50; 
zinc, 4.56;
germanium, 3.65;
arsenic, 2.30;
selenium, 3.34;
strontium, 2.87;
zirconium, 2.58;
barium, 2.18;
lanthanum, 1.10;
europium, 0.52;
platinum, 1.62; and
lead, 2.04.

Table~\ref{linetab} lists the 49~lines we have examined.
We adopt the \loggf\ values for Zn~\textsc{i} and \textsc{ii},
As~\textsc{i}, and Pt~\textsc{i} 
from \citet{roederer12c};
Ge~\textsc{i}, Sr~\textsc{ii}, and Ba~\textsc{ii}
from the NIST critical compilations \citep{klose02,fuhr09};
Se~\textsc{i} from \citet{morton00};
Zr~\textsc{ii} from \citet{ljung06} and \citet{malcheva06}, 
as discussed in the Appendix of \citet{roederer12b};
La~\textsc{ii} from \citet{lawler01a}; 
Eu~\textsc{ii} from \citet{lawler01b}; and
Pb~\textsc{i} from \citet{biemont00}.
Hyperfine splitting structure and isotope shifts are
considered for 
As~\textsc{i} \citep{roederer12c},
Ba~\textsc{ii} \citep{mcwilliam98}, 
La~\textsc{ii} \citep{ivans06},
Eu~\textsc{ii} \citep{ivans06}, and
Pb~\textsc{i} \citep{roederer12b}.
Uncertainties in the \loggf\ values for the elements of interest generally
do not
contribute significantly to the abundance uncertainties.
The Ge~\textsc{i}, As~\textsc{i}, and Se~\textsc{i} 
transitions used in this study
are all major branches \citep{meggers75}.
\citet{wiese80} estimate accuracies of 25\% 
($\leq$~0.10~dex) for the Ge~\textsc{i}
transition probabilities.
The As~\textsc{i} and Se~\textsc{i}
transition probabilities 
have stated uncertainties of 17\% or less
($\leq$~0.07~dex;
see references in Sections~3.3 and 3.4 of \citealt{roederer12c}).
These are relatively small when compared with other sources of 
uncertainty, but they are included in the random components of
uncertainty stated in Tables~\ref{abundtab1} and \ref{abundtab2}.

\begin{deluxetable*}{ccccccccccccccc}
\tablecaption{Mean Abundances for HD 2454, HD 16220, and HD 76932
\label{abundtab1}}
\tablewidth{0pt}
\tabletypesize{\scriptsize}
\tablehead{
\colhead{Element} &
\multicolumn{4}{c}{HD 2454} &
\colhead{} &
\multicolumn{4}{c}{HD 16220} &
\colhead{} &
\multicolumn{4}{c}{HD 76932} \\
\cline{2-5} \cline{7-10} \cline{12-15}
\colhead{or Ratio} &
\colhead{value} &
\colhead{$\sigma_{\rm random}$} &
\colhead{$\sigma_{\rm total}$} &
\colhead{$N$} &
\colhead{} &
\colhead{value} &
\colhead{$\sigma_{\rm random}$} &
\colhead{$\sigma_{\rm total}$} &
\colhead{$N_{\rm lines}$} &
\colhead{} &
\colhead{value} &
\colhead{$\sigma_{\rm random}$} &
\colhead{$\sigma_{\rm total}$} &
\colhead{$N$} }
\startdata
$\log \epsilon$\,(Fe~\textsc{i})  & $+$6.78 & 0.10    & 0.21    & 69 & & $+$6.94 & 0.14    & 0.24    & 51 & & $+$6.32 & 0.12    & 0.22    & 112 \\
$\log \epsilon$\,(Fe~\textsc{ii}) & $+$7.06 & 0.21    & 0.36    &  7 & & $+$7.09 & 0.22    & 0.36    &  7 & & $+$6.58 & 0.12    & 0.31    &  11 \\
$\log \epsilon$\,(Zn~\textsc{i})  & $+$3.97 & 0.07    & 0.20    &  3 & & $+$4.11 & 0.06    & 0.20    &  3 & & $+$3.71 & 0.06    & 0.20    &   4 \\
$\log \epsilon$\,(Zn~\textsc{ii}) & \nodata & \nodata & \nodata &  0 & & \nodata & \nodata & \nodata &  0 & & \nodata & \nodata & \nodata &   0 \\
$\log \epsilon$\,(Ge~\textsc{i})  & $+$2.70 & 0.14    & 0.24    &  1 & & $+$3.11 & 0.32    & 0.37    &  1 & & $+$2.40 & 0.14    & 0.24    &   1 \\
$\log \epsilon$\,(As~\textsc{i})  & $<+$2.9 & \nodata & \nodata &  2 & & $<+$3.1 & \nodata & \nodata &  1 & & $+$1.70 & 0.31    & 0.36    &   1 \\
$\log \epsilon$\,(Se~\textsc{i})  & $<+$3.3 & \nodata & \nodata &  1 & & $+$2.83 & 0.30    & 0.36    &  1 & & $+$2.41 & 0.16    & 0.25    &   2 \\
$\log \epsilon$\,(Sr~\textsc{ii}) & \nodata & \nodata & \nodata &  0 & & \nodata & \nodata & \nodata &  0 & & \nodata & \nodata & \nodata &   0 \\
$\log \epsilon$\,(Zr~\textsc{ii}) & $+$2.66 & 0.06    & 0.30    &  3 & & $+$2.33 & 0.05    & 0.29    &  6 & & $+$1.95 & 0.06    & 0.30    &  11 \\
$\log \epsilon$\,(Ba~\textsc{ii}) & $+$2.58 & 0.12    & 0.31    &  2 & & $+$1.99 & 0.33    & 0.44    &  3 & & $+$1.12 & 0.31    & 0.42    &   2 \\
$\log \epsilon$\,(La~\textsc{ii}) & $+$1.11 & 0.09    & 0.30    &  4 & & $+$0.82 & 0.06    & 0.30    &  4 & & $+$0.24 & 0.09    & 0.30    &   4 \\
$\log \epsilon$\,(Eu~\textsc{ii}) & $+$0.04 & 0.10    & 0.31    &  1 & & $+$0.14 & 0.10    & 0.31    &  1 & & $-$0.13 & 0.07    & 0.30    &   3 \\
$\log \epsilon$\,(Pt~\textsc{i})  & \nodata & \nodata & \nodata &  0 & & \nodata & \nodata & \nodata &  0 & & \nodata & \nodata & \nodata &   0 \\
$\log \epsilon$\,(Pb~\textsc{i})  & $+$2.08 & 0.10    & 0.21    &  1 & & \nodata & \nodata & \nodata &  0 & & $+$1.28 & 0.10    & 0.21    &   1 \\
\hline
~[Fe~\textsc{i}/H]               & $-$0.72 & 0.10    & \nodata & 69 & & $-$0.56 & 0.14    & \nodata & 51 & & $-$1.18 & 0.12    & \nodata & 112 \\
~[Fe~\textsc{ii}/H]              & $-$0.44 & 0.21    & \nodata &  7 & & $-$0.41 & 0.22    & \nodata &  7 & & $-$0.92 & 0.12    & \nodata &  11 \\
~[Zn~\textsc{i}/Fe~\textsc{i}]   & $+$0.13 & 0.12    & \nodata &  3 & & $+$0.11 & 0.15    & \nodata &  3 & & $+$0.33 & 0.13    & \nodata &   4 \\
~[Zn~\textsc{ii}/Fe~\textsc{ii}] & \nodata & \nodata & \nodata &  0 & & \nodata & \nodata & \nodata &  0 & & \nodata & \nodata & \nodata &   0 \\
~[Ge~\textsc{i}/Fe~\textsc{i}]   & $-$0.23 & 0.17    & \nodata &  1 & & $+$0.02 & 0.35    & \nodata &  1 & & $-$0.07 & 0.19    & \nodata &   1 \\
~[As~\textsc{i}/Fe~\textsc{i}]   & $<+$1.32& \nodata & \nodata &  2 & & $<+$1.36& \nodata & \nodata &  1 & & $+$0.58 & 0.33    & \nodata &   1 \\
~[Se~\textsc{i}/Fe~\textsc{i}]   & $<+$0.68& \nodata & \nodata &  1 & & $+$0.05 & 0.33    & \nodata &  1 & & $+$0.25 & 0.20    & \nodata &   2 \\
~[Sr~\textsc{ii}/Fe~\textsc{ii}] & \nodata & \nodata & \nodata &  0 & & \nodata & \nodata & \nodata &  0 & & \nodata & \nodata & \nodata &   0 \\
~[Zr~\textsc{ii}/Fe~\textsc{ii}] & $+$0.52 & 0.22    & \nodata &  3 & & $+$0.16 & 0.23    & \nodata &  6 & & $+$0.29 & 0.13    & \nodata &  11 \\
~[Ba~\textsc{ii}/Fe~\textsc{ii}] & $+$0.84 & 0.24    & \nodata &  2 & & $+$0.22 & 0.40    & \nodata &  3 & & $-$0.14 & 0.33    & \nodata &   2 \\
~[La~\textsc{ii}/Fe~\textsc{ii}] & $+$0.45 & 0.23    & \nodata &  4 & & $+$0.13 & 0.23    & \nodata &  4 & & $+$0.06 & 0.15    & \nodata &   4 \\
~[Eu~\textsc{ii}/Fe~\textsc{ii}] & $-$0.04 & 0.23    & \nodata &  1 & & $+$0.03 & 0.24    & \nodata &  1 & & $+$0.27 & 0.14    & \nodata &   3 \\
~[Pt~\textsc{i}/Fe~\textsc{i}]   & \nodata & \nodata & \nodata &  0 & & \nodata & \nodata & \nodata &  0 & & \nodata & \nodata & \nodata &   0 \\
~[Pb~\textsc{i}/Fe~\textsc{i}]   & $+$0.76 & 0.14    & \nodata &  1 & & \nodata & \nodata & \nodata &  0 & & $+$0.42 & 0.16    & \nodata &   1 \\
\enddata
\end{deluxetable*}

\begin{deluxetable*}{ccccccccccccccc}
\tablecaption{Mean Abundances for HD 94028, HD 107113, and HD 140283
\label{abundtab2}}
\tablewidth{0pt}
\tabletypesize{\scriptsize}
\tablehead{
\colhead{Element} &
\multicolumn{4}{c}{HD 94028} &
\colhead{} &
\multicolumn{4}{c}{HD 107113} &
\colhead{} &
\multicolumn{4}{c}{HD 140283} \\
\cline{2-5} \cline{7-10} \cline{12-15}
\colhead{or Ratio} &
\colhead{value} &
\colhead{$\sigma_{\rm random}$} &
\colhead{$\sigma_{\rm total}$} &
\colhead{$N$} &
\colhead{} &
\colhead{value} &
\colhead{$\sigma_{\rm random}$} &
\colhead{$\sigma_{\rm total}$} &
\colhead{$N_{\rm lines}$} &
\colhead{} &
\colhead{value} &
\colhead{$\sigma_{\rm random}$} &
\colhead{$\sigma_{\rm total}$} &
\colhead{$N$} }
\startdata
$\log \epsilon$\,(Fe~\textsc{i})  & $+$5.75 & 0.09   & 0.21   & 44 & & $+$6.71 & 0.08    & 0.21    & 58 & & $+$4.79 & 0.05    & 0.20    & 109 \\
$\log \epsilon$\,(Fe~\textsc{ii}) & $+$5.88 & 0.05   & 0.29   &  6 & & $+$6.97 & 0.07    & 0.30    &  5 & & $+$4.88 & 0.06    & 0.30    &  28 \\
$\log \epsilon$\,(Zn~\textsc{i})  & $+$3.09 & 0.06   & 0.20   &  3 & & $+$3.83 & 0.06    & 0.20    &  3 & & $+$2.13 & 0.07    & 0.20    &   2 \\
$\log \epsilon$\,(Zn~\textsc{ii}) & $+$3.26 & 0.10   & 0.31   &  1 & & \nodata & \nodata & \nodata &  0 & & $+$2.27 & 0.10    & 0.31    &   1 \\
$\log \epsilon$\,(Ge~\textsc{i})  & $+$1.56 & 0.17   & 0.26   &  3 & & $+$2.54 & 0.22    & 0.29    &  1 & & $<+$0.6 & \nodata & \nodata &   4 \\
$\log \epsilon$\,(As~\textsc{i})  & $+$1.21 & 0.21   & 0.28   &  1 & & $+$1.87 & 0.31    & 0.36    &  1 & & $+$0.09 & 0.21    & 0.28    &   1 \\
$\log \epsilon$\,(Se~\textsc{i})  & $+$2.06 & 0.17   & 0.25   &  3 & & $+$2.77 & 0.21    & 0.28    &  1 & & $+$0.60 & 0.24    & 0.31    &   3 \\
$\log \epsilon$\,(Sr~\textsc{ii}) & $+$1.51 & 0.07   & 0.30   &  2 & & \nodata & \nodata & \nodata &  0 & & $+$0.09 & 0.07    & 0.30    &   2 \\
$\log \epsilon$\,(Zr~\textsc{ii}) & $+$1.46 & 0.07   & 0.30   &  6 & & $+$2.06 & 0.10    & 0.31    &  1 & & $-$0.15 & 0.08    & 0.30    &   3 \\
$\log \epsilon$\,(Ba~\textsc{ii}) & $+$0.68 & 0.13   & 0.32   &  2 & & $+$1.82 & 0.14    & 0.32    &  2 & & $-$1.37 & 0.10    & 0.31    &   3 \\
$\log \epsilon$\,(La~\textsc{ii}) & $-$0.28 & 0.05   & 0.29   &  4 & & $+$0.67 & 0.05    & 0.29    &  4 & & $<-$1.4 & \nodata & \nodata &   4 \\
$\log \epsilon$\,(Eu~\textsc{ii}) & $-$0.88 & 0.10   & 0.31   &  1 & & $<+$0.4 & \nodata & \nodata &  1 & & $<-$1.7 & \nodata & \nodata &   3 \\
$\log \epsilon$\,(Pt~\textsc{i})  & $+$0.20 & 0.10   & 0.21   &  1 & & \nodata & \nodata & \nodata &  0 & & $<-$0.6 & \nodata & \nodata &   1 \\
$\log \epsilon$\,(Pb~\textsc{i})  & $<+$1.3 & \nodata&\nodata &  1 & & \nodata & \nodata & \nodata &  0 & & $<+$0.7 & \nodata & \nodata &   1 \\
\hline
~[Fe~\textsc{i}/H]               & $-$1.75 & 0.09   & \nodata & 44 & & $-$0.79 & 0.08    & \nodata & 58 & & $-$2.71 & 0.05    & \nodata & 109 \\
~[Fe~\textsc{ii}/H]              & $-$1.62 & 0.05   & \nodata &  6 & & $-$0.53 & 0.07    & \nodata &  5 & & $-$2.62 & 0.06    & \nodata &  28 \\
~[Zn~\textsc{i}/Fe~\textsc{i}]   & $+$0.28 & 0.11   & \nodata &  3 & & $+$0.06 & 0.10    & \nodata &  3 & & $+$0.28 & 0.09    & \nodata &   2 \\
~[Zn~\textsc{ii}/Fe~\textsc{ii}] & $+$0.32 & 0.11   & \nodata &  1 & & \nodata & \nodata & \nodata &  0 & & $+$0.33 & 0.12    & \nodata &   1 \\
~[Ge~\textsc{i}/Fe~\textsc{i}]   & $-$0.34 & 0.19   & \nodata &  3 & & $-$0.32 & 0.24    & \nodata &  1 & & $<-$0.34& \nodata & \nodata &   4 \\
~[As~\textsc{i}/Fe~\textsc{i}]   & $+$0.66 & 0.23   & \nodata &  1 & & $+$0.36 & 0.32    & \nodata &  1 & & $+$0.50 & 0.22    & \nodata &   1 \\
~[Se~\textsc{i}/Fe~\textsc{i}]   & $+$0.47 & 0.19   & \nodata &  3 & & $+$0.22 & 0.23    & \nodata &  1 & & $-$0.03 & 0.25    & \nodata &   3 \\
~[Sr~\textsc{ii}/Fe~\textsc{ii}] & $+$0.26 & 0.09   & \nodata &  2 & & \nodata & \nodata & \nodata &  0 & & $-$0.16 & 0.09    & \nodata &   2 \\
~[Zr~\textsc{ii}/Fe~\textsc{ii}] & $+$0.50 & 0.09   & \nodata &  6 & & $+$0.01 & 0.12    & \nodata &  1 & & $-$0.11 & 0.10    & \nodata &   3 \\
~[Ba~\textsc{ii}/Fe~\textsc{ii}] & $+$0.12 & 0.14   & \nodata &  2 & & $+$0.17 & 0.16    & \nodata &  2 & & $-$0.93 & 0.12    & \nodata &   3 \\
~[La~\textsc{ii}/Fe~\textsc{ii}] & $+$0.24 & 0.07   & \nodata &  4 & & $+$0.10 & 0.09    & \nodata &  4 & & $<+$0.12& \nodata & \nodata &   4 \\
~[Eu~\textsc{ii}/Fe~\textsc{ii}] & $+$0.22 & 0.11   & \nodata &  1 & & $<+$0.41& \nodata & \nodata &  1 & & $<+$0.40& \nodata & \nodata &   3 \\
~[Pt~\textsc{i}/Fe~\textsc{i}]   & $+$0.33 & 0.13   & \nodata &  1 & & \nodata & \nodata & \nodata &  0 & & $<+$0.49& \nodata & \nodata &   1 \\
~[Pb~\textsc{i}/Fe~\textsc{i}]   & $<+$1.01& \nodata& \nodata &  1 & & \nodata & \nodata & \nodata &  0 & & $<+$1.37& \nodata & \nodata &   1 \\
\enddata
\end{deluxetable*}

We adopt an \rpro\ isotopic mix \citep{sneden08}
for the metal-poor stars
and a Solar isotopic mix \citep{bohlke05} for the more metal-rich stars
with a relatively large content of \spro\ material.
We do not report abundances for strontium, platinum, or lead in most stars
in the sample, unfortunately, 
because the lines are either undetected or 
extremely blended.
The limited wavelength coverage available for \hde\ does not
cover the strongest lines of Eu~\textsc{ii}, so we can
only derive an upper limit from the non-detection of the
intrinsically weak 
Eu~\textsc{ii} 6645\,\AA\ transition.

We compute uncertainties assuming a minimum measurement (random)
uncertainty of 0.10~dex, 
as estimated from unblended Fe~\textsc{i} lines.
This accounts for uncertainty in the continuum placement,
line fitting, and any mild blending features.
Lines marked in Table~\ref{linetab} with ``:'' indicate we have adopted
a minimum measurement uncertainty of 0.20~dex, and lines
marked with  ``::'' indicate we have adopted
a minimum measurement uncertainty of 0.30~dex.
These lines are more blended, and
we have had greater difficulty identifying
the local continuum.
Following \citet{roederer12c}, we assume a systematic
uncertainty of 0.19~dex for absorption lines of neutral atoms
and 0.29~dex for absorption lines of ions,
which accounts for uncertainties in the model atmosphere parameters.
The total uncertainty is the quadrature sum of the random and
systematic uncertainties, as given in Tables~\ref{abundtab1}
and \ref{abundtab2}.
Most abundance ratios constructed from species in the same ionization
state should be fairly robust against uncertainties in the
model atmosphere parameters.
The combined random uncertainties should adequately represent the
uncertainty in these cases, so no total uncertainties are given.

We derive germanium abundances from several lines, including the
Ge~\textsc{i} 3039\,\AA\ line that has been used previously
\citep{sneden98,sneden03,cowan02,cowan05} 
to study germanium in metal-poor stars.
Unfortunately we have not been able to detect both the UV lines
near 2000\,\AA\ and the 3039\,\AA\ line simultaneously,
so we cannot verify that these abundance indicators 
yield consistent results.
The subgiants studied in this study and \citet{roederer12c}
yield [Ge/Fe] ratios or upper limits that are higher than the
samples of giants studied by \citet{cowan05} and \citet{roederer12b}.
The subgiant sample occupies higher metallicities, 
so this offset does not necessarily indicate a systematic difference.
This issue could be examined with new observations
of both sets of Ge~\textsc{i} lines in the same star.

\citet{roederer12c} estimated that substantial fractions 
($\approx$~10\%--90\%) of neutral arsenic and selenium 
are present in the line-forming layers of the atmosphere of
\hdg.
Both elements have relatively high first ionization potentials
(FIPs),
9.79~eV and 9.75~eV, respectively.
Germanium has a somewhat lower FIP, 7.90~eV.
Our calculations reveal that $\approx$~3\%--9\% of the germanium
in the line-forming layers is neutral; 
while small,
this is a substantially higher percentage of Ge~\textsc{i} than,
e.g., Fe~\textsc{i} 
expected in the line-forming
layers of these warm subgiants.
The lower levels of the Ge~\textsc{i} transitions used in our study
comprise the main population reservoirs ($>$~98\%) for Ge~\textsc{i}
at the temperatures found in these atmospheres,
although the fraction of all germanium atoms in 
a given level never exceeds 1\%--2\%.
Our calculations assume that the level populations and 
ionization fractions can be described by Boltzmann and Saha
distributions in local thermodynamical equilibrium (LTE).
We caution that our LTE calculations may
underestimate the total germanium abundance
due to the minority species Ge~\textsc{i} being driven out of LTE by 
overionization.
Estimating the possible magnitude of this effect 
is beyond the scope of the present study, although 
we would expect the non-LTE corrections to be similar
for all stars in our sample.
Non-LTE calculations of the germanium line formation
in these stars would be most welcome.

\citet{mashonkina12} have estimated non-LTE corrections
to the abundances derived from the 
Pb~\textsc{i} 4057\,\AA\ and Eu~\textsc{ii} 4129\,\AA\ lines
in the sample of stars examined by \citet{roederer10c}.
Neutral lead atoms are no more than a couple percent of the total
fraction of lead in metal-poor subgiants or turnoff stars,
so the corrections may be substantial.
\citeauthor{mashonkina12}\ calculate non-LTE corrections of
approximately $+$0.25 to $+$0.40~dex for the 
Pb~\textsc{i} 4057\,\AA\ line in stars like 
those in our sample.
Their calculated non-LTE corrections for 
the Eu~\textsc{ii} 4129\,\AA\ line are small,
typically $+$0.05~dex.
We do not include these corrections to the values presented
in Tables~\ref{abundtab1} and \ref{abundtab2}, 
but we caution that the lead abundances listed may be
underestimated by a factor of 2 or so.

For optical wavelengths of interest to abundance 
studies of late-type stars,
the continuous opacity is dominated by 
bound-free (bf) absorption of the H$^{-}$ ion.
Rayleigh scattering 
and bf absorption from atomic H
also contribute to a lesser degree.
At UV wavelengths in stars of high metallicity, 
bf absorption from neutral metals 
also becomes a significant source of continuous opacity.
Our calculations indicate that
the opacity is dominated by H$^{-}_{\rm bf}$ absorption
in these stars
at most wavelengths of interest to the present study.
H~\textsc{i}$_{\rm bf}$ also contributes significantly
at wavelengths short of the Balmer jump.
Bf absorption from the 
Mg~\textsc{i} 
$3s3p$ $^{3}P^{\rm o}$ term
and the
Al~\textsc{i}
$3s^{2}3p$ $^{2}P^{\rm o}$ term
occurs short of 2515\,\AA\ and 2076\,\AA, respectively.
The contributions to the total continuous opacity from these metals
are comparable to---but do not dominate---the contributions from H 
only in the most metal-rich stars 
([Fe/H]~$\approx -$0.5) in our sample.
In these stars,
abundance ratios derived from lines in our UV spectra near 2000\,\AA\
could be referenced to either H or Al,
since these elements both contribute roughly
equally to the total continuous opacity.
In the more metal-poor stars,
H still dominates the sources of continuous opacity near 2000\,\AA.
We underscore the fact that most of the lines of interest of 
Ge~\textsc{i}, As~\textsc{i}, and Se~\textsc{i}
occur within a narrow wavelength range, 
so abundance ratios among these elements
should be quite robust.

Our derived mean abundances 
are listed in Tables~\ref{abundtab1} and \ref{abundtab2}.
Both $\log \epsilon$ abundances and [X/Fe] ratios are given.
Recall that abundances or ratios denoted with the ionization state
represent the total abundance of that element as derived from
absorption lines of the species indicated.
Note that the [Zn/Fe] ratios derived from both neutral and 
ionized zinc are in agreement in \hdd\ and \hdf, as 
\citet{roederer12c} also found for \hdg.

\section{Results and Discussion}
\label{results}

\subsection{The Nature of Our Sample}

\begin{figure}
\begin{center}
\includegraphics[angle=0,width=3.4in]{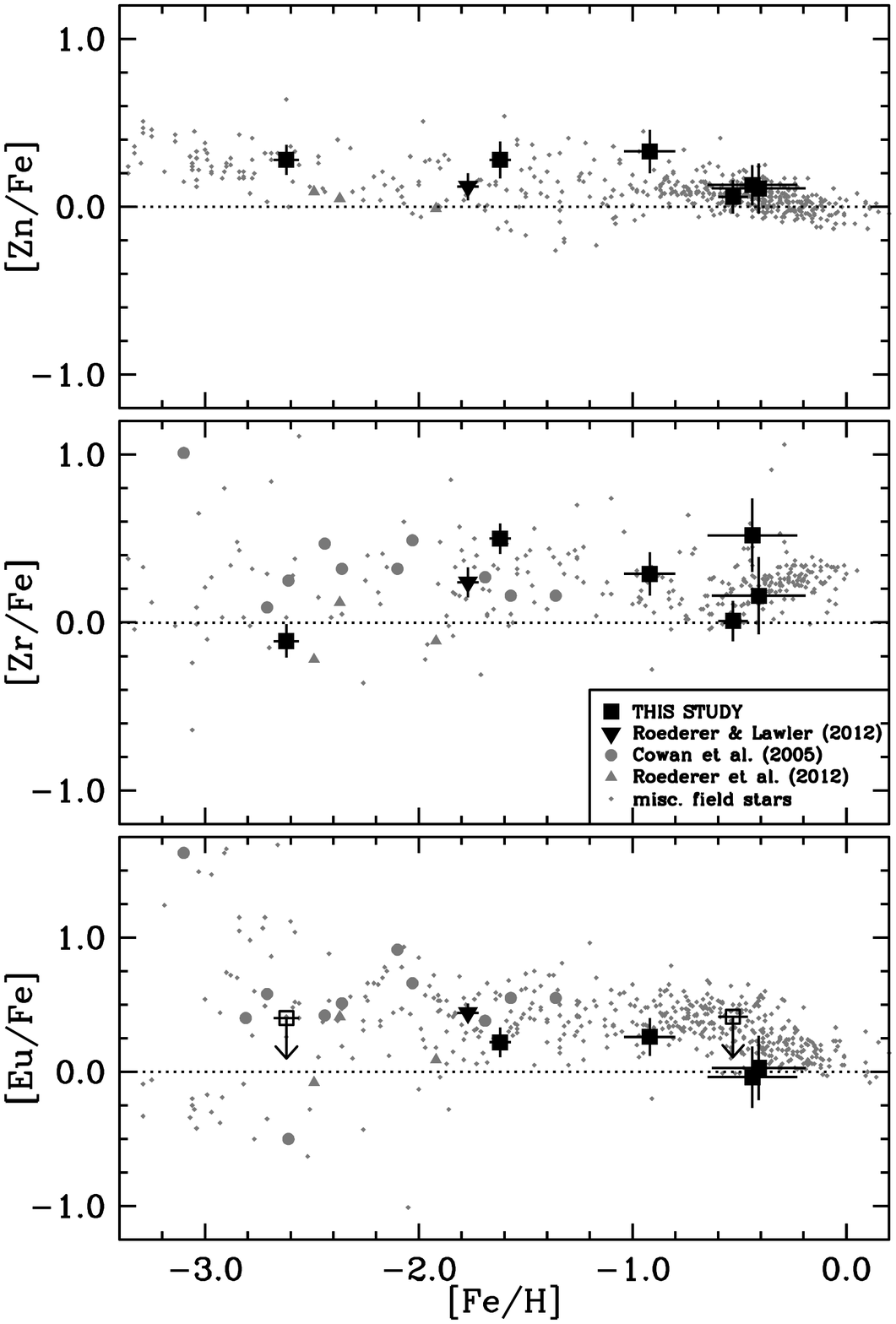}
\end{center}
\caption{
\label{znplot}
The [Zn/Fe], [Zr/Fe], and [Eu/Fe] ratios as a function of [Fe/H].
Stars examined in this study are marked by the large black squares.
The other star where germanium, arsenic, and selenium have been studied,
\hdg, is marked by the large black triangle.
Other stars where germanium has been studied are marked by
medium-sized gray circles \citep{cowan05} or medium-sized gray triangles
\citep{roederer12b}.
These studies included several stars
that had been examined previously by \citet{sneden98,sneden03}
and \citet{cowan02}.
Filled symbols denote detections, and open symbols denote upper limits.
Abundances of field stars derived by various surveys
are taken from \citet{burris00}, \citet{fulbright00}, \citet{reddy03,reddy06}, 
\citet{cayrel04}, \citet{francois07}, and \citet{roederer10c}.
The dotted horizontal lines indicate the Solar ratios.
 }
\end{figure}

The stars in our sample span a wide range of ages, metallicities,
and Galactic stellar populations.
The three most metal-rich stars in our sample, 
\hda, \hdb, and \hde, have
overall metallicities [Fe/H]~$\approx -$0.5.
These stars have kinematics consistent with membership in the 
thin disc, are approximately 2--4~Gyr old, and have only
mildly enhanced [$\alpha$/Fe] ratios (where $\alpha$ stands
for O, Mg, Si, Ca, or Ti), 
$+$0.05~$\leq$~[$\alpha$/Fe]~$\leq +$0.15
(e.g., \citealt{edvardsson93,venn04,bensby05,holmberg09,casagrande11}).
\hdc\ is more metal poor, [Fe/H]~$= -$0.9,
is a probable member of the thick disc,
has an intermediate age of approximately 7--11~Gyr, and has
significant $\alpha$-enhancement, [$\alpha$/Fe]~$= +$0.3
(e.g., \citealt{edvardsson93,venn04,
nissen10,casagrande11,schuster12}).
The three most metal-poor stars, \hdd, \hdf, and \hdg, 
have metallicities spanning $-$2.6~$\leq$~[Fe/H]~$\leq -$1.6.
These stars have kinematics consistent with the thick disc or halo,
have ages of 12~Gyr or more, and have enhanced
[$\alpha$/Fe] ratios, $+$0.3~$\leq$~[$\alpha$/Fe]~$\leq +$0.4
(e.g., \citealt{fulbright00,venn04,holmberg09,
casagrande11,roederer12c}).

In Figure~\ref{znplot}, we compare the [Zn/Fe], [Zr/Fe], and 
[Eu/Fe] ratios for the stars in our sample 
(black squares and triangle) with previous
surveys of the Galactic halo, thick, and thin discs
(small diamonds).
We also highlight stars with previous [Ge/Fe] determinations
(gray circles and triangles).
When necessary, we correct the \loggf\ scales to 
those used in the present study.\footnote{
Large surveys concerned with the chemical evolution of the thin
disc find that the mean [Zr/Fe] ratio converges to the Solar
ratio at Solar metallicity (e.g., \citealt{edvardsson93,reddy03}).
In our Figure~\ref{znplot}, however,
the Zr~\textsc{ii} abundances for stars near Solar metallicity,
mostly derived by \citeauthor{reddy03},
appear to converge toward [Zr/Fe]~$= +$0.3 at [Fe/H]~$=$~0.0.
\citeauthor{reddy03}\ used one Zr~\textsc{ii} line, 5112\,\AA, and they adopted
its \loggf\ value from \citet{biemont81} and \citet{bogdanovich96},
which is 0.26~dex higher than the value presented by \citet{ljung06}.
The discrepancy stems from the smaller experimental branching
fraction measured by \citeauthor{ljung06}\ relative to the 
values available previously.
The \loggf\ value given by \citeauthor{ljung06}\ yields a
Zr~\textsc{ii} abundance derived from the
the 5112\,\AA\ line in the Solar photosphere that is in better
agreement with other lines, and we prefer this value.
}
The two most metal-rich stars in our sample have [Eu/Fe] ratios
consistent with the Solar ratio, which is on the low end of 
the field star distribution near [Fe/H]~$= -$0.4. 
One of them, \hda, is a known barium dwarf star
and will be discussed separately in Section~\ref{hd2454}.
Overall these stars
fall well within the ranges of [Zn/Fe], [Zr/Fe], and [Eu/Fe]
for field stars at similar metallicity.
This demonstrates that the stars of interest for the present study
are not outliers according to their chemistry.

\subsection{Germanium Trends with Metallicity}

\begin{figure}
\begin{center}
\includegraphics[angle=0,width=3.4in]{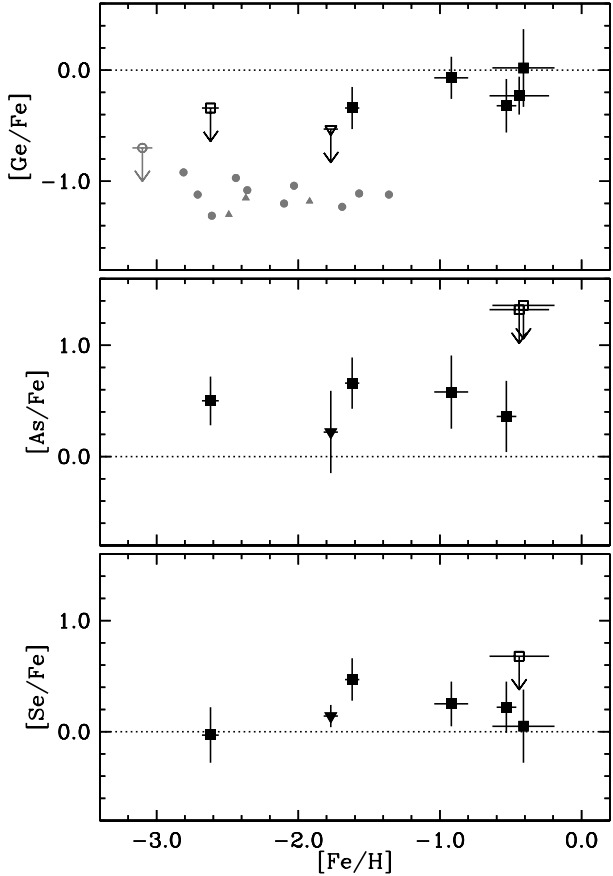}
\end{center}
\caption{
\label{geplot}
The [Ge/Fe], [As/Fe], and [Se/Fe] ratios as a function of [Fe/H].
Symbols are the same as in Figure~\ref{znplot}.
Filled symbols denote detections, and open symbols denote upper limits.
 }
\end{figure}

\begin{figure}
\begin{center}
\includegraphics[angle=0,width=3.4in]{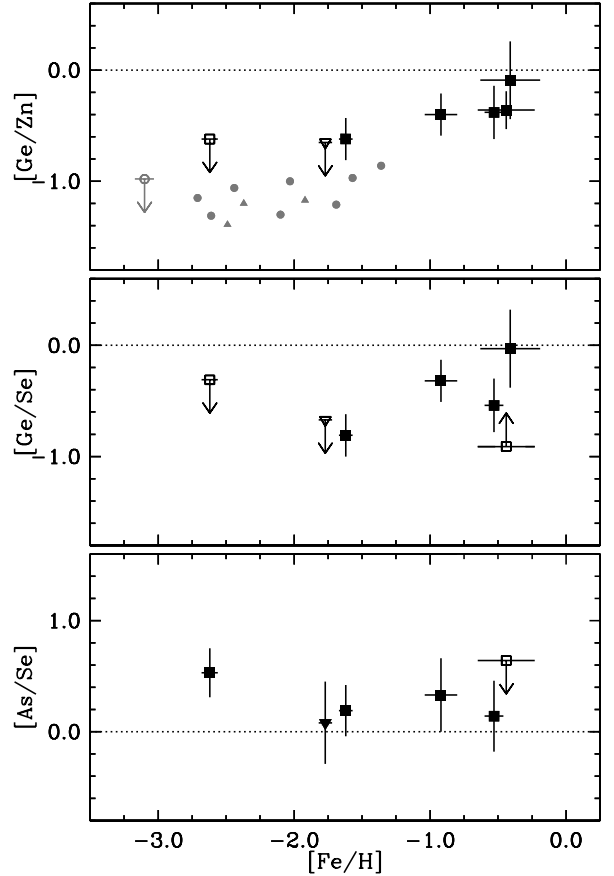}
\end{center}
\caption{
\label{seplot}
The [Ge/Zn], [Ge/Se], and [As/Se] ratios as a function of [Fe/H].
Symbols are the same as in Figure~\ref{znplot}.
Filled symbols denote detections, and open symbols denote limits.
The upward arrow in the middle panel denotes a lower limit on [Ge/Se]
from an upper limit on the selenium abundance.
 }
\end{figure}

Figures~\ref{geplot} and \ref{seplot} indicate that the
[Ge/Fe] and [Ge/Zn] ratios increase with increasing metallicity.
The increase in [Ge/Fe] begins at least by [Fe/H]~$= -$1.6,
and the four stars in our sample with [Fe/H]~$> -$1.0 show 
[Ge/Fe] ratios close to the Solar ratio.
Other stars with [Fe/H]~$= -$1.6
show low [Ge/Fe] ratios that are consistent with those found in 
the class of \rpro\ rich standard stars, whose members include
\mbox{BD$+$17~3248} \citep{cowan02},
\mbox{CS~22892--052} \citep{sneden03},
\mbox{CS~31082--001} \citep{hill02},
\mbox{HD~115444} \citep{westin00}, and
\mbox{HD~221170} \citep{ivans06}.
Previous
observations of metal-poor stars show that [Ge/Fe]
does not correlate with [Eu/Fe], 
a common tracer of the main component of the \rpro,
indicating that germanium is not produced under the same
conditions that give rise to \rpro\ nucleosynthesis. 
This result was initially found by \citet{cowan05},
and more recent observations reaffirm it \citep{roederer12b}.

The overall increase and scatter
in [Ge/Fe] echoes the increase and scatter in [La/Eu]
found in many field stars over the same metallicity range,
as shown in Figure~\ref{laeuplot}.
The lowest metallicity stars and some even at metallicity
[Fe/H]~$= -$1.4 show a low [La/Eu] ratio that is consistent with
the [La/Eu] ratio found in the \rpro\ rich standard stars
(shaded band in Figure~\ref{laeuplot}).
A range of [La/Eu] ratios is found at most metallicities,
but the upturn in [La/Eu] toward the Solar ratio
with increasing metallicity generally is attributed
to increasing amounts of \spro\ material present in the ISM
when these stars were born
(e.g., \citealt{simmerer04}).
It is reasonable to conclude that the increase in [Ge/Fe] ratios
may also be due to increasing amounts of \spro\ material
present in the ISM.
We dedicate the next sections to investigating the nature of the 
\spro\ that could be responsible for the germanium production.

\subsection{The Barium Star HD~2454}
\label{hd2454}

\citet{tomkin89} identified 
\hda\ (\mbox{HR~107}), which is in our sample,
as the first example of a barium dwarf star.
Such stars may be the dwarf analogs of the classical barium
giants, Population~I stars in binary systems
with strong lines of Ba~\textsc{ii}, Sr~\textsc{ii}, 
C$_{2}$, CH, and CN. 
These stars may be formed by 
accreting mass from a more massive companion that passed through
the AGB phase of evolution.

A recent study of the heavy element abundance pattern in
\hda\ by \citet{allen06} reveals substantial overabundances of 
elements at all three \spro\ peaks, as well as
a mild carbon overabundance,
supporting this interpretation.
The \spro\ pattern in this star (either their abundances or ours)
can be reasonably fit by the \spro\ abundance predictions
of \citet{husti09} and \citet{bisterzo10}
for an AGB star with an initial main sequence mass of 
1.2~$\lesssim M/M_{\odot} \lesssim$~3.0, 
metallicity [Fe/H]~$\approx -$0.5, and a fairly standard
$^{13}$C pocket efficiency.
All but the stars of lowest mass in this range 
will evolve in less than $\approx$~2~Gyr,
which is less than the inferred age of the observed main sequence star,
$\approx$~2.0--2.5~Gyr \citep{bensby05,casagrande11}.

We find that
the [Zr/Fe], [Ba/Fe], and [La/Fe] ratios in \hda\ 
are enhanced by factors of 2--4
relative to these ratios in \hdb, which has a nearly identical
metallicity and [Eu/Fe] ratio as \hda.
(The [Pb/Fe] ratio in \hda\ is also enhanced; 
we could not derive [Pb/Fe] in \hdb.)
The [Ge/Fe] ratios are identical within the uncertainties in 
\hda\ and \hdb, however, suggesting that the nucleosynthesis
process responsible for the enhancements at the three \spro\ peaks
in the AGB companion
did not produce any detectable germanium.
In this scenario, the germanium enrichment was present
in the birth cloud of \hda, while most of the enrichment 
of zirconium, barium, and lanthanum occurred 
via mass transfer from a binary companion that passed through
the AGB phase.

\subsection{Nucleosynthesis of Germanium}

Our data indicate that the nucleosynthesis process responsible
for the increasing [Ge/Fe] ratios should produce
substantial germanium, 
relatively small amounts of zinc, arsenic, and selenium
(see Section~\ref{asse}), and little material
with $A \gtrsim$~90.
If we attribute the 
enrichment of elements at the \spro\ peaks in \hda\ to the 
main component of the \spro, then the overall upturn
in [Ge/Fe] 
could be the result of the weak component of the \spro.
This is consistent with longstanding interpretations of the
origin of the
A~$\lesssim$~90 portion of the Solar system \spro\ distribution
(e.g., \citealt{kappeler89}; \citealt{raiteri91}),
where 80\% of the germanium is estimated to have originated
in the weak component of the \spro\ operating in massive
($M \sim$~25~$M_{\odot}$) stars
(e.g., \citealt{pignatari10}).

\begin{figure}
\begin{center}
\includegraphics[angle=0,width=3.4in]{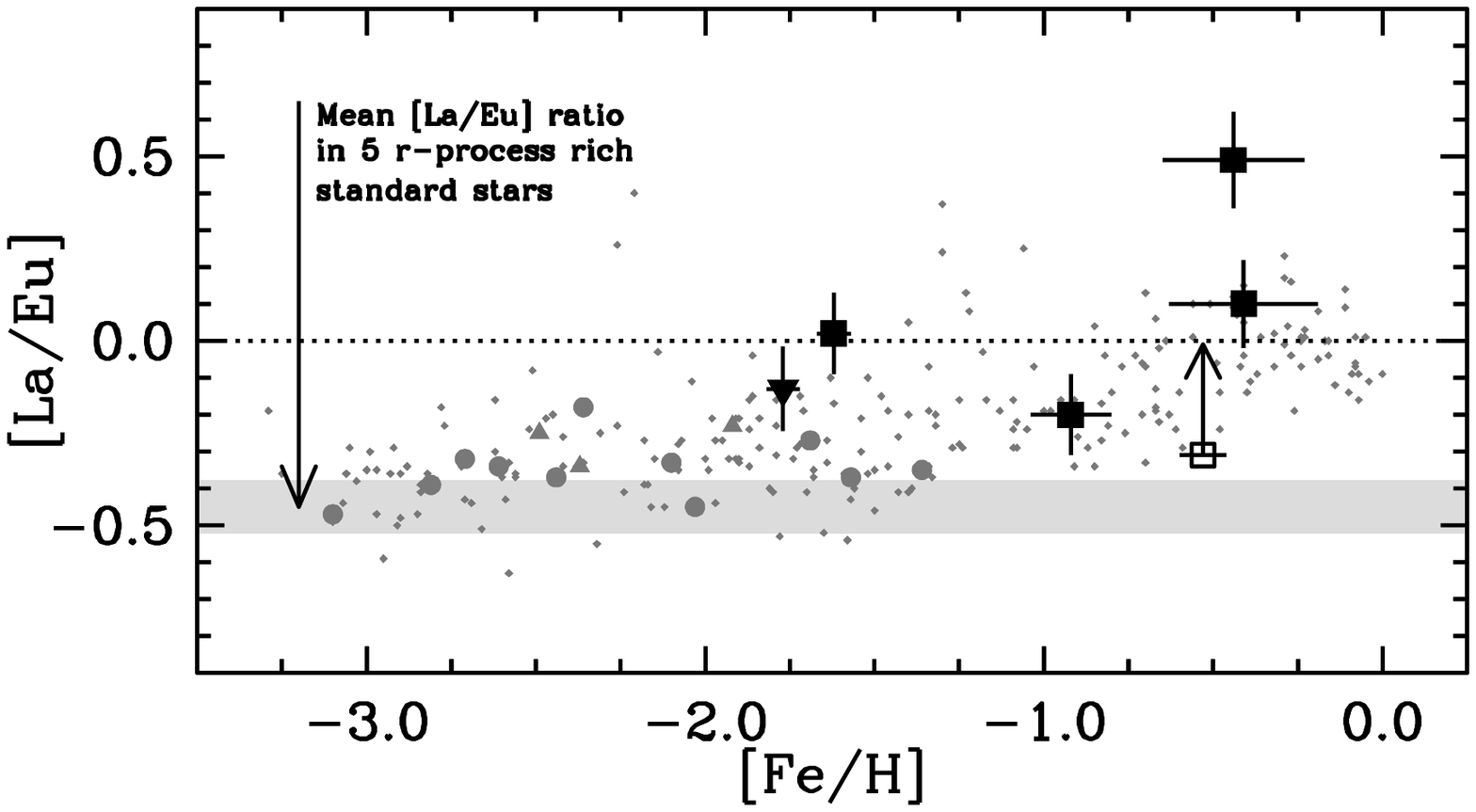}
\end{center}
\caption{
\label{laeuplot}
The [La/Eu] ratio as a function of [Fe/H].
Symbols are the same as in Figure~\ref{znplot}.
Filled symbols denote detections, and open symbols denote limits.
The upward arrow denotes a lower limit on [La/Eu]
from an upper limit on the europium abundance.
The gray shaded band represents the mean [La/Eu] ratio found in 
5 \rpro\ rich standard stars, as analyzed by \citet{sneden09}, 
$\langle$[La/Eu]$\rangle = -$0.45~$\pm$~0.07.
The field sample includes additional stars from
\citet{simmerer04} and their reanalysis of stars examined by \citet{woolf95}.
 }
\end{figure}

The germanium in the 
more metal-poor stars seems to have been produced 
by a different mechanism. 
The germanium in these stars may have been produced
by previous stellar generations under conditions of
nuclear statistical equilibrium
(e.g., \citealt{wanajo11}) or charged-particle reactions
(e.g., \citealt{frohlich06}).
Perhaps
the ISM in the birth environments of most metal-poor stars
was not widely enriched by products of the weak \spro.
Alternatively, the \spro\ that was in operation in 
massive, low-metallicity stars 
may have been fundamentally different
than that in metal-rich stars.

\subsection{Arsenic and Selenium Trends with Metallicity}

Figures~\ref{geplot} and \ref{seplot} show the evolution 
of the [As/Fe], [Se/Fe], and [As/Se] ratios 
as a function of metallicity.
Although there are only 7~stars available at present
to trace the evolution of these elements,
a few points can be inferred from these figures.
There appears to be relatively little
evolution in either the [As/Fe] or [Se/Fe] ratios
from $-$2.6~$<$~[Fe/H]~$< -$0.4.
The mean [As/Fe] ratio is
$+$0.51~$\pm$~0.08 ($\sigma =$~0.18), and the
mean [Se/Fe] ratio is 
$+$0.19~$\pm$~0.07 ($\sigma =$~0.18).
Unsurprisingly, the [As/Se] trend is also relatively flat. 
In contrast, the [Ge/Se] ratio
suggests an upward trend with increasing metallicity.
At low metallicity, [Fe/H]~$< -$0.9, the [As/Fe] ratios
are enhanced similarly to [Eu/Fe].
[Se/Fe] is less enhanced but still super-Solar,
except possibly in \hdf.
Our metal-poor sample does not include any stars with extreme
values of [Eu/Fe], 
so we cannot investigate trends between arsenic or selenium
and europium as \citet{cowan05} did for germanium.
We could only derive upper limits on arsenic and selenium in \hda, 
the barium dwarf star, since the
lines are too blended to be reliable abundance indicators.

\subsection{Nucleosynthesis of Arsenic and Selenium}
\label{asse}

Our observations indicate [Ge/Se] is trending toward the Solar ratio
at high metallicity, while [As/Se] remains Solar or super-Solar.
Neither [As/Fe] nor [Se/Fe] appear to be affected by 
the weak \spro\ in the way that [Ge/Fe] is.
This result is consistent with
the relatively small fraction of arsenic and selenium
in the Solar system attributed to the \spro.
The rapidly-rotating massive star models of \citet{frischknecht12}
predict that \spro\ production of germanium
should be accompanied by production of arsenic and selenium
by similar factors.
While the weak \spro\ may
contribute small amounts of arsenic and selenium
that are undetected amidst a relatively large primary abundance,
our data exclude it as a major source of the
arsenic and selenium in the metal-rich stars.

\hdb, \hdc, \hdd, \hde, and \hdg\ all have [La/Eu] ratios
enhanced significantly above the mean [La/Eu] found in the 
\rpro\ standard stars, as shown in Figure~\ref{laeuplot}.
This indicates
that material produced by the main component of the \spro\
was incorporated into these stars.
The intermediate mass
AGB models of \citet{karakas09} that cover a range of metallicities predict
approximately Solar [Ge/Se] ratios, and the low mass
models predict [Ge/Se]~$\approx -$0.15 to $-$0.20. 
This is also the case at very low metallicity;
see the models in \citet{lugaro12}.
These models also predict that [As/Se] should be
sub-Solar by a factor of $\approx$~2
(A.\ Karakas, 2012, private communication).
The relatively constant
and super-Solar [As/Fe], [Se/Fe], and [As/Se] ratios
indicate that the main component of the \spro\
also
contributed little to these elements in our stellar sample.

Simulations of \rpro\ nucleosynthesis predict that conditions
leading to the production
of these elements should be different than the conditions leading
to the main component of the \rpro\
(e.g., \citealt{kratz07,farouqi09}).
As with germanium in low metallicity stars, 
explosive nucleosynthesis in 
nuclear statistical equilibrium (e.g., \citealt{wanajo11})
or a charged-particle process (e.g., \citealt{frohlich06})
are likely candidates, 
but these data cannot affirm this specifically.
Detection of arsenic and selenium in 
stars with [Fe/H]~$< -$2.0
and low [Ge/Fe] and [La/Eu] ratios would 
help diagnose this matter.

\section{Conclusions}
\label{conclusions}

We present a study of the
chemical evolution of germanium, arsenic, and selenium
in metal-poor stars.
Of these three elements, only germanium abundances 
have been studied previously in multiple metal-poor stars,
and that only for stars with [Fe/H]~$\leq -$1.4 
\citep{sneden98,sneden03,cowan02,cowan05,roederer12b}.
Using archive space-based UV
spectra obtained with STIS on board the \textit{HST}
and ground-based optical spectra from several 
observatories, we are able to study 
abundances of these elements 
in seven stars spanning metallicities 
$-$2.6~$\leq$~[Fe/H]~$\leq -$0.4.
We also derive abundances of zinc, strontium,
zirconium, barium, lanthanum, europium, platinum, and lead
to inform our interpretation.

The stars in our sample appear to be normal chemical representatives
of the halo, thick, and thin discs.
Previous observations indicate that
germanium tracks iron 
at low metallicity \citep{cowan05}.
Our new observations indicate that [Ge/Fe] ratios increase
at higher metallicities,
presumably due to
production by the weak component of the \spro.
[As/Fe] and [Se/Fe] show no such increase, suggesting that
their dominant production mechanism may be the
same at both low and high metallicity.
Explosive nucleosynthesis or charged-particle reactions
are likely mechanisms for arsenic and selenium production,
though our data are agnostic about this point.
Our data imply that
the main component of the \spro\ 
has contributed very little to
the germanium, arsenic, or selenium in our stellar sample.

These results are not in conflict with 
theoretical predictions 
for the origin of
germanium, arsenic, and selenium in the Solar system.
Considering the scarce abundance data on these elements
available beyond the Solar abundance distribution,
we find this agreement encouraging.

Finally, we underscore that our conclusions 
have been drawn from abundances in a very small number of stars.
Studies of arsenic and selenium in metal-poor stars
are reliant on access to high resolution, high S/N
UV spectra attainable only with \textit{HST}.
This has been, and will remain, the limiting factor in 
such studies for the foreseeable future.

\acknowledgments

I am especially grateful to the many observers whose data
have made this study possible.
I thank A.\ Karakas and M.\ Pignatari for
helpful conversations on \spro\ nucleosynthesis and
the referee, Charles Cowley, for a careful reading of
the manuscript.
This research has made use of NASA's
Astrophysics Data System Bibliographic Services,
the arXiv pre-print server operated by Cornell University,
the SIMBAD and VizieR databases hosted by the
Strasbourg Astronomical Data Center,
the Atomic Spectra Database hosted by
the National Institute of Standards and Technology, 
the Mikulski Archive at the Space Telescope Science Institute,
the ESO Science Archive Facility,
and the Keck Observatory Archive.
IRAF is distributed by the National Optical Astronomy Observatories,
which are operated by the Association of Universities for Research
in Astronomy (AURA), Inc., under cooperative agreement with the National
Science Foundation.
This research is supported in part by NASA through a grant 
to Program GO-12268
from the Space Telescope Science Institute, which is operated
by AURA under NASA contract NAS~5-26555.
I.U.R.\ is supported by the Carnegie Institution of Washington 
through the Carnegie Observatories Fellowship.

{\it Facilities:} 
\facility{ESO:3.6m (HARPS)},
\facility{HST (STIS)},
\facility{Keck:I (HIRES)},
\facility{Smith (Tull)},
\facility{VLT:Kueyen (UVES)}

\end{document}